\documentclass[prd,amsmath,showpacs,preprintnumbers,
               floatfix,twocolumn,letterpaper]{revtex4}

\usepackage{graphicx}
\usepackage{slashed}

\newcommand{\nc}{\newcommand}

\nc{\eref}[1]{~(\ref{#1})}
\nc{\etal}{\emph{~et al.}}
\nc{\ecite}[1]{~\cite{#1}}
\nc{\efig}[1]{~Fig.~\ref{#1}}
\nc{\etable}[1]{~Table~\ref{#1}}

\nc{\cor}{\langle qq \rangle}
\nc{\longcor}{\langle q(x) q(0) \rangle}

\nc{\tr}{{\rm{Tr}}}
\nc{\retr}{{\rm{Re\, tr}}}
\nc{\dg}{\dagger}
\nc{\hm}{\hat{\mu}}
\nc{\hn}{\hat{\nu}}
\nc{\eq}[1]{~(\ref{#1})}
\nc{\Fmunu}{F_{\mu\nu}}
\nc{\Dmu}{{\mathcal D}_\mu}
\nc{\Dnu}{{\mathcal D}_\nu}
\nc{\fundDmu}{D_\mu}
\nc{\fundDnu}{D_\nu}
\nc{\scD}{{\mathcal D}}

\nc{\FSq}{\Fmunu^2(x)}
\nc{\DmuFSq}{(\Dmu\Fmunu(x))^2}
\nc{\DnuFSq}{(\Dnu\Fmunu(x))^2}
\nc{\DmuSqFSq}{(\Dmu^2\Fmunu(x))^2}
\nc{\DnuSqFSq}{(\Dnu^2\Fmunu(x))^2}
\nc{\DmuSqFDnuSqF}{\Dmu^2\Fmunu(x)\Dnu^2\Fmunu(x)}
\nc{\DmuDnuFSq}{(\Dmu\Dnu\Fmunu(x))^2}
\nc{\FSqSq}{\Fmunu^4(x)}

\begin{document}

\preprint{ADP-08-02/T661}

\title{Impact of Dynamical Fermions on QCD Vacuum Structure}

\author{Peter J. Moran} \affiliation{Special Research Centre for the
  Subatomic Structure of Matter (CSSM), Department of Physics,
  University of Adelaide 5005, Australia} \author{Derek B. Leinweber}
\affiliation{Special Research Centre for the Subatomic Structure of
  Matter (CSSM), Department of Physics, University of Adelaide 5005,
  Australia}


\begin{abstract}
  We examine how dynamical fermions affect both the UV and infrared
  structure of the QCD vacuum.  We consider large $28^3 \times 96$
  lattices from the MILC collaboration, using a gluonic definition of
  the topological charge density, founded on a new over-improved
  stout-link smearing algorithm.  The algorithm reproduces established
  results from the overlap formalism and is designed to preserve
  nontrivial topological objects including instantons.  At short
  distances we focus on the topological charge correlator, $\langle
  q(x) q(0) \rangle$, where negative values at small $x$ reveal a
  sign-alternating layered structure to the topological-charge density
  of the QCD vacuum.  We find that the magnitudes of the negative dip
  in the $\langle q(x)q(0) \rangle$ correlator and the positive
  $\langle q(0)^2 \rangle$ contact term are both increased with the
  introduction of dynamical fermion degrees of freedom.  This is in
  accord with expectations based on charge renormalization and the
  vanishing of the topological susceptibility in the chiral limit.  At
  large distances we examine the extent to which instanton-like
  objects are found on the lattice, and how their distributions vary
  between quenched and dynamical gauge fields.  We show that dynamical
  gauge fields contain more instanton-like objects with an average
  size greater than in the quenched vacuum.  Finally, we directly
  visualize the topological charge density in order to investigate the
  effects of dynamical sea-quark degrees of freedom on topology.
\end{abstract}

\maketitle

\section{Introduction}

The study of QCD vacuum structure is one area of research where
lattice simulations provide access to otherwise unaccessible
information. By generating typical vacuum gauge field configurations
we are able to directly investigate their complex structure. Further,
by varying the simulation parameters we can assess how different
physical phenomena contribute to the structure of the vacuum. In
particular, we are able to examine the effect of dynamical sea-quarks
on the QCD vacuum.

A study of QCD vacuum structure at different scales usually requires
the use of a filtering procedure. By removing the short-range UV
fluctuations, one can probe the long-distance structural features of
the vacuum. Typical UV filtering methods include
cooling~\cite{cool1,cool2,cool3}, APE~\cite{ape1,ape2}, and improved
APE smearing~\cite{Bonnet:2001rc}, HYP smearing~\cite{hyp} and more
recently, stout-link smearing~\cite{Morningstar} and LOG
smearing~\cite{Durr:2007cy}.  More recently, the truncation of a
spectral representation of operators using eigenmodes of the Overlap
Dirac operator has been explored as a method to remove UV fluctuations
\cite{Horvath:2002yn,Ilgenfritz:2007xu,Bruckmann:2007ww,newilgenfritz}.

Instantons are believed to be an essential component of the
long-distance physics of the QCD vacuum. The problem with most cooling
and smearing algorithms is that they destroy the instantons in the
vacuum. In particular, standard EXP, or stout-link, smearing suffers
from this problem. In order to overcome this we employ a form of
over-improved stout-link smearing~\cite{Moran:2007nc,Moran:2008ra},
described briefly in Sec.~\ref{sec:overimprovedstoutlinksmearing}.

One facet of recent topological studies of vacuum structure is the
topological charge density correlator, or Euclidean 2-point
correlation function
\begin{equation}
  \cor \equiv \longcor
  \label{eq:cor} \, ,
\end{equation}
which is the integrand of the topological susceptibility
\begin{equation}
  \chi \equiv \frac{\langle Q^2 \rangle}{V} = \int\!d^4x\, \longcor \,,
  \label{eqn:topsus}
\end{equation}
where $V$ is the 4-volume.

In Sec.~\ref{sec:topologicalchargedensitycorrelator}, we examine the
shapes of the $\cor$ correlator, using both over-improved stout-link
smearing and 3-loop, $\mathcal{O}(a^4)$ improved cooling.  We use the
gluonic definition of the topological charge density
\begin{equation}
  q(x) = \frac{g^2}{32 \pi^2} \epsilon_{\mu\nu\rho\sigma} \tr [
  F_{\mu\nu}(x) F_{\rho\sigma}(x) ] \,,
\end{equation}
where
\begin{equation}
  Q = \sum_x q(x) \,,
\end{equation}
and $F_{\mu\nu}(x)$ is a 3-loop improved field strength
tensor\ecite{BilsonThompson:2002jk}.
Of particular interest is the extent to which a purely gluonic
definition of the topological charge density can revel a negative
$\cor$ correlator. We then proceed to investigate how the correlator
is affected by the inclusion of dynamical sea-quarks.

In Sec.~\ref{sec:instantonlikeobjects} we probe the infrared structure
of the full dynamical QCD vacuum. We examine the extent to which
instantons are found in the QCD vacuum and how the extra dynamical
fermion degrees of freedom affect their distribution throughout the
vacuum.

Finally, in Sec.~\ref{sec:topologicalchargedensity} we directly
visualize the topological charge density in order to investigate the
effects of dynamical sea-quark degrees of freedom on topology.
Conclusions are summarized in Sec.~\ref{sec:conclusion}.
\section{Over-Improved Stout-Link Smearing}
\label{sec:overimprovedstoutlinksmearing}

Extended smoothing on any QCD gauge field will reveal the presence of
smooth topological objects that are approximations to the classical
Euclidean instanton solution. A single instanton has the gauge
potential~\cite{Belavin:1975fg}
\begin{equation}
  A_\mu(x) = \frac{x^2}{x^2 + \rho_{\rm inst}^2} \left( \frac{i}{g} \right)
  \partial_\mu (S) \, S^{-1} \, ,
  \label{eqn:instantonsoln}
\end{equation}
where
\begin{equation}
  S \equiv \frac{x_4 \pm i \, \vec{x} \cdot \vec{\sigma} }{ \sqrt{x^2}} \,,
\end{equation}
for instantons and anti-instantons.

Due to their correlations with low-lying Dirac eigenmodes these
locally self-dual instanton-like objects are of obvious physical
interest. Unfortunately it is well realized that standard smoothing
procedures will often destroy instantons in the gauge field.

For this investigation we have chosen to employ over-improved
stout-link smearing\ecite{Moran:2008ra}, and will now present a brief
summary of the algorithm.
Complete details and an investigation of the algorithm's performance
are provided in Ref.~\cite{Moran:2008ra}.
We also present results using 3-loop improved
cooling\ecite{BilsonThompson:2002jk}, and defer the reader
elsewhere\ecite{Bonnet:2001rc,BilsonThompson:2001ca,BilsonThompson:2002jk,deForcrand:1995qq}
for a discussion of highly improved actions.

Standard stout-link smearing is closely related to the Wilson action
\begin{equation}
  S_{W}(x) = \beta \sum_{\scriptstyle \nu \atop \scriptstyle \mu \ne \nu}
  \frac{1}{3} {\retr} \left [ 1 - U_{\mu}(x) \Sigma_{\mu \nu}(x) \right ] \,,
  \label{eqn:localwilsonaction}
\end{equation}
where
\begin{equation}\label{eqn:wilsonstaples}
  \begin{split}
    \Sigma_{\mu \nu}(x) &= U_{\nu}(x+\hm) \, U^{\dg}_{\mu}(x+\hn) \, U^\dg_{\nu}(x)\\
    &\quad +U^{\dg}_{\nu}(x+\hm-\hn) \, U^{\dg}_{\mu}(x-\hn) \,
    U_{\nu}(x-\hn) \,,
  \end{split}
\end{equation}
to calculate the nearest neighbor contributions. However, it is
possible to use any reasonable combination of links connected to the
original link $U_\mu(x)$. We can analyze the effect of the Wilson
action on an instanton by substituting the instanton solution into the
Taylor expansion of $S_W(x)$~\cite{GarciaPerez:1993ki,Moran:2008ra}
%
%
%
\begin{equation}
  S^{inst}_W = \frac{8 \pi^2}{g^2} \left[ 1 - \frac{1}{5} \left( \frac{a}{\rho_{\rm inst}} \right)^{\!\!2} - \frac{1}{70}\left( \frac{a}{\rho_{\rm inst}} \right)^{\!\!4} \right] \, .
  \label{eqn:instwilsaction}
\end{equation}
If one plots this expression as a function of the instanton size
$\rho_{\rm inst}$ one sees that the slope of the action is necessarily
always positive.  Denoting the action associated with a single
classical instanton as $S_0$, we note that the ideal action for
preserving instantons in the gauge fields would have a flat curve at
$S/S_0=1$, as in the continuum.

In order to tame the lattice discretisation errors, improved actions
have been developed, such as the Symanzik\ecite{symanzik}
$O(a^2)$-improved action
\begin{equation}
  \begin{split}
    S_{S} & = \beta \sum_x \sum_{\mu > \nu} \bigg[
    \frac{5}{3} ( 1 - P_{\mu\nu}(x) ) \\
    &- \frac{1}{12} \big( ( 1 - R_{\mu\nu}(x) ) + ( 1 - R_{\nu\mu}(x)
    ) \big) \bigg] \,,
  \end{split}
  \label{eqn:symanzikaction}
\end{equation}
where we have $R_{\mu\nu}$ and $R_{\nu\mu}$ to denote the different
possible orientations of the rectangular loop, and $P_{\mu\nu}$
denotes the plaquette.

We modify this action, following Perez\etal\ecite{GarciaPerez:1993ki},
by the inclusion of a new $\epsilon$ parameter
\begin{equation}
  \begin{split}
    S(\epsilon) & = \beta \sum_x \sum_{\mu > \nu} \bigg[
    \frac{5-2\epsilon}{3} ( 1 - P_{\mu\nu}(x) ) \\
    &- \frac{1-\epsilon}{12} \big( ( 1 - R_{\mu\nu}(x) ) + ( 1 -
    R_{\nu\mu}(x) ) \big) \bigg] \,, \\
  \end{split}
  \label{eqn:overimpaction}
\end{equation}
where for $\epsilon=1$ we have the standard Wilson action, and for
$\epsilon=0$ we have the Symanzik improved action. So by varying
$\epsilon$ between $0$ and $1$ we can effectively select how much
improvement is included.

Alternatively we can set $\epsilon < 0$ and over-improve the action.
The advantage of a negative $\epsilon$ value is that the leading order
errors become positive upon substitution of the instanton solution
\begin{equation}
  S^{inst}(\epsilon) = \frac{8 \pi^2}{g^2} \left[ 1 
    - \frac{\epsilon}{5} \left(\frac{a}{\rho_{\rm inst}}\right)^{\!\!2}
    + \frac{14\epsilon-17}{210} \left(\frac{a}{\rho_{\rm inst}}\right)^{\!\!4} \right] \,.
\end{equation}
The curve of $S/S_0$ is similarly affected. One sees that for a value
of $\epsilon = -0.25$ the curve is mostly flat and will preserve
instantons objects on the lattice with size $\rho_{\rm inst}/a >
1.5$~\cite{Moran:2008ra}.

Standard stout-link smearing, using an isotropic smearing parameter
$\rho_{\rm sm}$, involves a simultaneous update of all links on the
lattice. Each link is replaced by a smeared link $\tilde{U}_\mu(x)$
\begin{equation}
  \tilde{U}_\mu(x) = \mathrm{exp}(i Q_\mu(x) ) \, U_\mu(x) \,,
  \label{eqn:stoutlinksmearedlink}
\end{equation}
where
\begin{align}
  Q_\mu(x) & = \frac{i}{2}(\Omega_\mu^\dagger(x) - \Omega_\mu(x)) \notag \\
  & \quad - \frac{i}{6}\tr(\Omega_\mu^\dagger(x) - \Omega_\mu(x)) \,,
\end{align}
with
\begin{equation}
  \Omega_\mu(x) = \rho_{\rm sm}\,\sum\{1\times1{\rm\ loops\ involving\ }U_\mu(x) \} \,.
\end{equation}
The over-improvement parameter $\epsilon$ is introduced into the
smearing process by replacing the combination of links in
$\Omega_\mu(x)$ with
\begin{equation}
  \begin{split}
    \label{eq:stoutcmux}
    \Omega_\mu&(x) = \rho_{\rm sm}\,\sum\bigg\{ \frac{5-2\epsilon}{3}
    ({\rm 1\times1\ loops\ involving\ }U_\mu(x) ) \\
    &-\frac{1-\epsilon}{12} ({\rm 1\times2+2\times1\ loops\ involving\
    }U_\mu(x) ) \bigg\} \,.
  \end{split}
\end{equation}
Note that both forward and backward horizontally orientated rectangles
are included in the $2\times1$ loops, such that $\Omega_\mu(x)$
resembles the local action.
With this extended link path, we take $\rho_{\rm sm} = 0.06$, smaller
than the standard value of 0.1.

\section{Topological Charge Density Correlator}
\label{sec:topologicalchargedensitycorrelator}

\subsection{Quenched QCD}

The Euclidean 2-point correlation function for the topological charge
$q(x)$
\begin{equation}
  \cor \equiv \longcor \, ,
\end{equation}
also known as the topological charge density correlator, is negative
for any $x > 0$\ecite{Seiler:1987ig,Seiler:2001je}.  This follows
simply from reflection positivity\ecite{Horvath:2005cv}. Given that
the correlator must have a positive contact term $\langle q^2(0)
\rangle_x$, the correlator necessarily has the form
\begin{equation}
  \longcor = A \delta(x) - f(x) \, ,
\end{equation}
where $f(x)$ is positive for all $x$.

Recent studies\ecite{Horvath:2005cv,Ilgenfritz:2007xu} of vacuum
structure using the overlap defined topological charge density have
demonstrated the negativity of the topological charge density
correlator in lattice simulations.
The negative behavior of the $\cor$ correlator suggests a
sign-alternating layered structure to the topological charge density
of the topological charge density correlator.

Hasenfratz\ecite{Hasenfratz} has also examined the shape of the
2-point correlator using APE smearing. For APE smearing the magnitude
of the negative dip was far less than the current overlap
results\ecite{Horvath:2005cv,Ilgenfritz:2007xu}. We now attempt to
obtain a negative $\cor$ correlator similar to the overlap results
using the new over-improved stout-link smearing algorithm and 3-loop
improved gluonic operators for the action and topological charge
densities\ecite{BilsonThompson:2002jk}. In order to more closely
compare with the overlap results we use quenched gauge fields for the
initial calculation.

The gauge fields were generated by the MILC
collaboration\ecite{milc1,milc2} using a Tadpole and Symanzik improved
gauge action with $1\times1 + 1\times2 + 1\times1\times1$ terms in the
quenched case and an Asqtad staggered dynamical fermion action for the
$2+1$ flavors of dynamical quarks. The lattice spacing for all three
types of gauge fields is $a = 0.09$~fm. For the specifics of how the
gauge fields were generated see Refs.\ecite{milc1,milc2}. Simulation
parameters are summarized in\etable{table:gaugefields}.
\begin{table}
  \caption{\label{table:gaugefields} The gauge fields used in this
    study. The lattices were generated by the MILC
    collaboration\ecite{milc1,milc2}}
  \begin{ruledtabular}
    \begin{tabular}{lccccr}

      Size & $\beta$ & $a$ & Bare Quark Masses \\
      $28^3 \times 96$ & 8.40 & $0.086$ fm & $\infty$ \\
      $28^3 \times 96$ & 7.09 & $0.086$ fm & $14.0\,\mathrm{MeV}, 67.8\,\mathrm{MeV}$ \\
      $28^3 \times 96$ & 7.11 & $0.086$ fm & $27.1\,\mathrm{MeV}, 67.8\,\mathrm{MeV}$
    \end{tabular}
  \end{ruledtabular}
\end{table}

\begin{figure}
  
  \includegraphics[angle=90,width=0.45\textwidth]{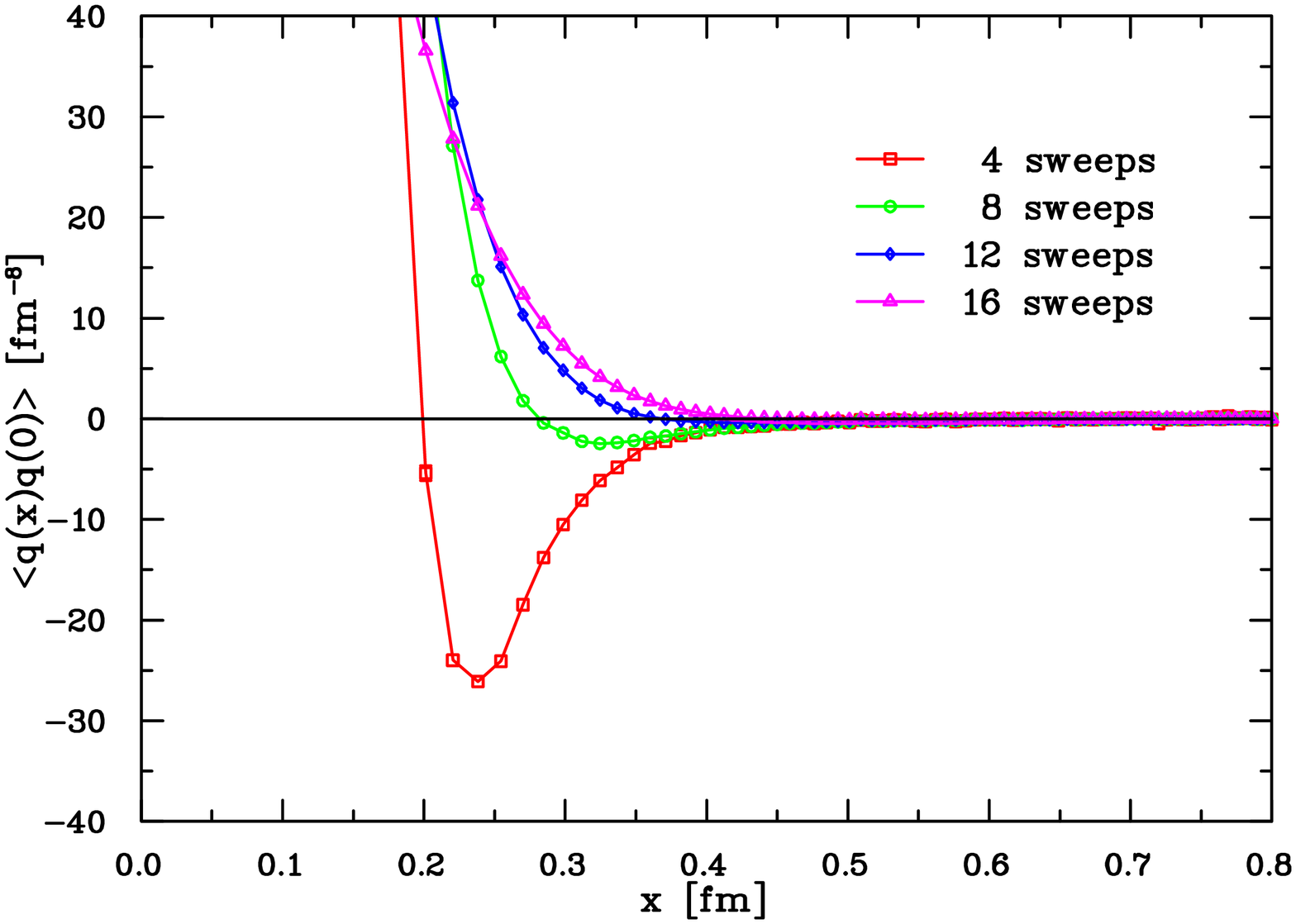}\\[10pt]
  \includegraphics[angle=90,width=0.45\textwidth]{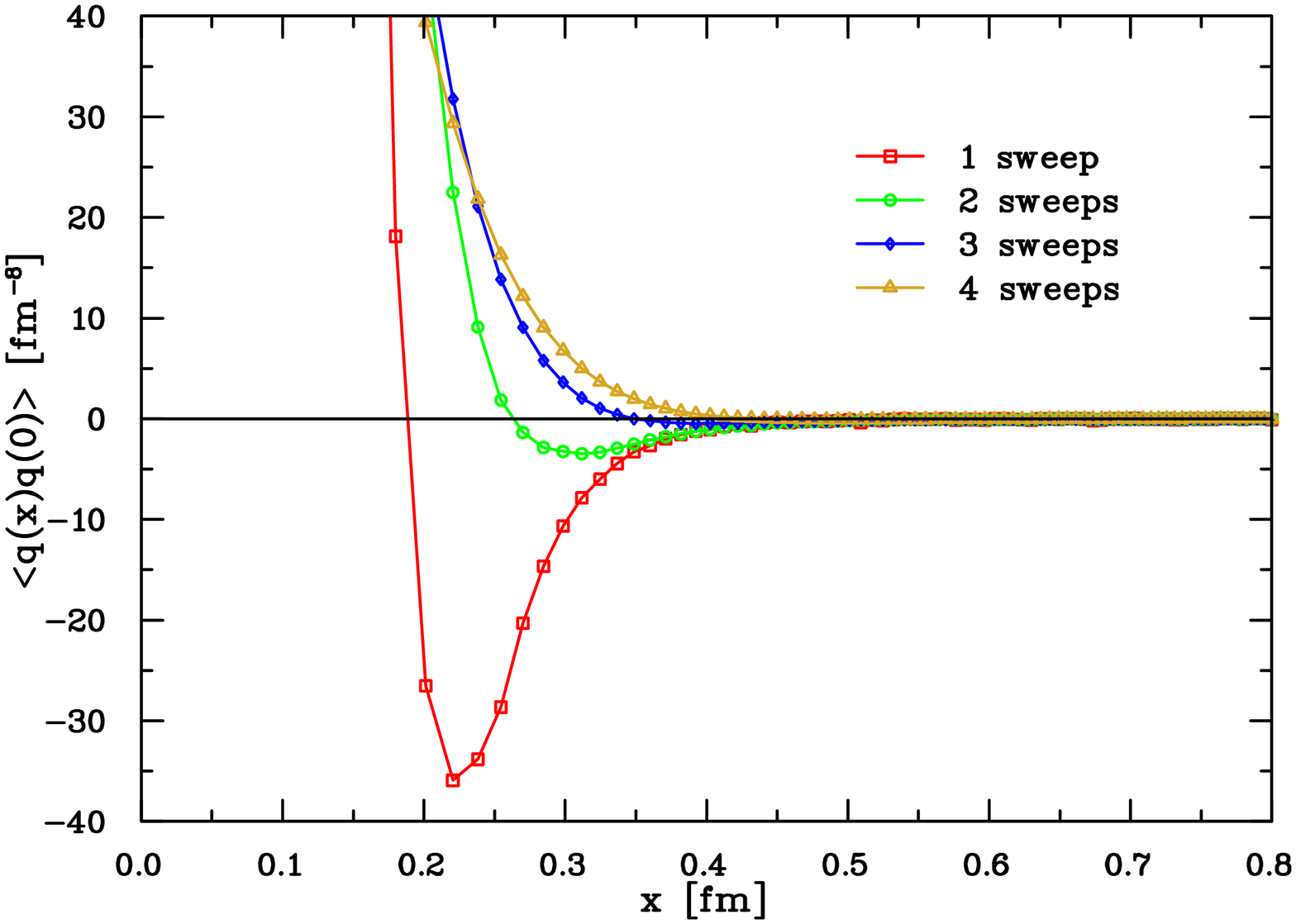}
  
  \caption{The $\cor$ correlator for a variable number of smoothing
    sweeps calculated on the quenched gauge fields. The top graph
    shows the $\cor$ correlator for over-improved stout-link smearing
    smearing, and the lower graph for 3-loop improved cooling. Using
    four sweeps of smearing and one sweep of cooling it is possible to
    obtain a correlator similar to the recent overlap
    results\ecite{Horvath:2005cv,Ilgenfritz:2007xu}. As more UV
    fluctuations are removed in successive smoothing iterations the
    correlator flattens off. Errors were calculated using the
    jackknife procedure but are too small to see.}
  \label{fig:qqcalibn}
\end{figure}

We begin by investigating how the shape of the correlator changes as a
function of the number of smearing sweeps. The results of our
calculations are presented in\efig{fig:qqcalibn}.
We see that for a small number of sweeps it is possible to generate a
negative correlator, but that the negative behavior is largely
suppressed after about 10 sweeps and absent after around 40 sweeps.

The curves also have a positive core whose minimal size is limited by
the lattice spacing and the combination of loops in the smoothing
procedure. As more of the short-distance UV fluctuations are removed
by the smoothing algorithm there is a dampening effect on the
topological correlator.

\begin{figure}

  \includegraphics[angle=0,width=0.45\textwidth]{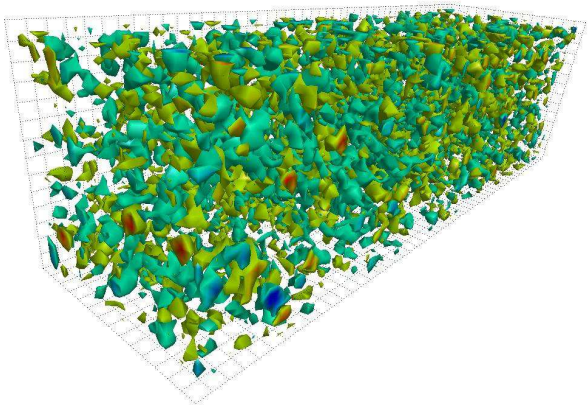}
  
  \caption{The topological charge density of a quenched gauge field
    after four sweeps of smearing.  Note the sign-alternating
    structure in the charge density.}
  \label{fig:qx.qnchd.4sweeps}
\end{figure}

One sweep of highly improved cooling gives a $\cor$ correlator of
similar proportions to the overlap defined
correlator\ecite{Horvath:2005cv,Ilgenfritz:2007xu}, whilst for
over-improved smearing, four sweeps reveals a similar curve.  A
quantitative comparison of the overlap correlator is performed in
Ref.~\cite{newilgenfritz}.  The topological charge density of a
quenched field after 4 sweeps of smearing is presented in
Fig.~\ref{fig:qx.qnchd.4sweeps}.
There is considerable structure within the gauge field.  A
sign-alternating structure in the topological charge density is also
apparent.


\subsection{QCD with Dynamical Fermions}

We now investigate the effects of dynamical sea-quark degrees of
freedom on topology. In quenched QCD the
Witten-Veneziano\ecite{Witten,Veneziano} formula gives a relation
between the topological susceptibility and the mass of the
$\eta^\prime$ meson\ecite{Seiler:2001je}
\begin{equation}
  \chi^{qu} = \frac{m_{\eta^\prime}^2 F_\pi^2}{2N_f} \,.
\end{equation}
However, in the full dynamical case the topological susceptibility
should vanish in the chiral limit\ecite{Hart:2001pj,Seiler:2001je}
\begin{equation}
  \chi^{dyn} = \frac{ f_\pi^2 m_\pi^2}{2 N_f} + \mathcal{O}(m_\pi^4) \,.
  \label{eqn:fullchi}
\end{equation}

Of course, a vanishing topological susceptibility puts no restraints
on how the function $\longcor$ should change with the addition of
dynamical sea-quarks, it only requires that the integral of
Eq.\eref{eqn:topsus} vanish.

Including dynamical sea-quarks in the QCD action renormalizes the
coupling constant. In order to maintain the same lattice spacing
across quenched and dynamical gauge fields, one finds that the
coupling parameter, $g$, must increase and hence $\beta \sim 1/g^2$
must be smaller for the dynamical fields.  In the QCD action
formulated in Euclidean space, $\beta$ appears as a factor governing
the width of the probability distribution.  When generating gauge
fields, smaller $\beta$ values permit greater fluctuations in the
gauge links. The increased fluctuations can give rise to non-trivial
field fluctuations, which will reveal itself in lattice simulations
through a greater mean-square density $\langle q^2(0) \rangle_x$.

\begin{figure}
  
  \includegraphics[angle=90,width=0.45\textwidth]{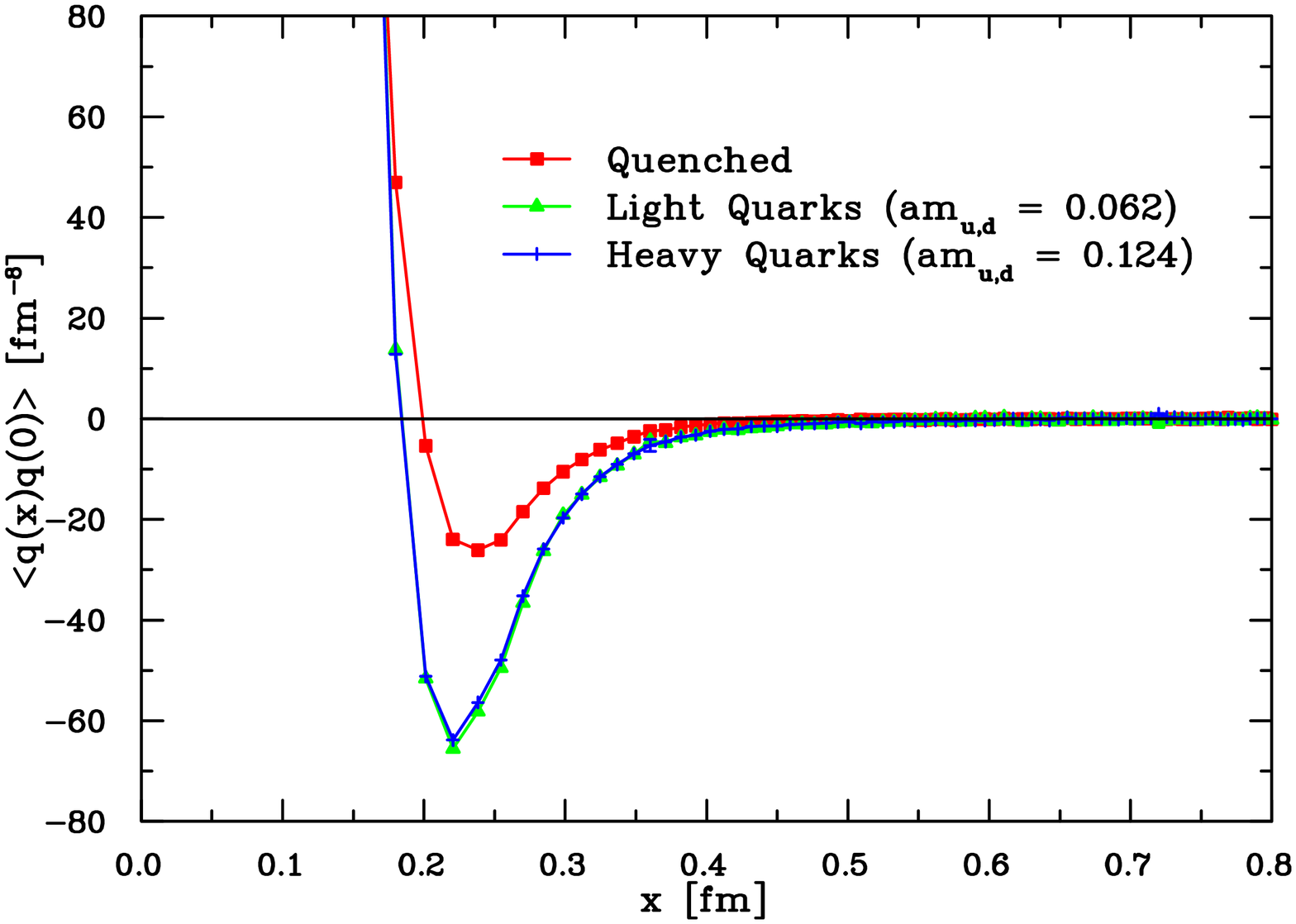}\\[10pt]
  \includegraphics[angle=90,width=0.45\textwidth]{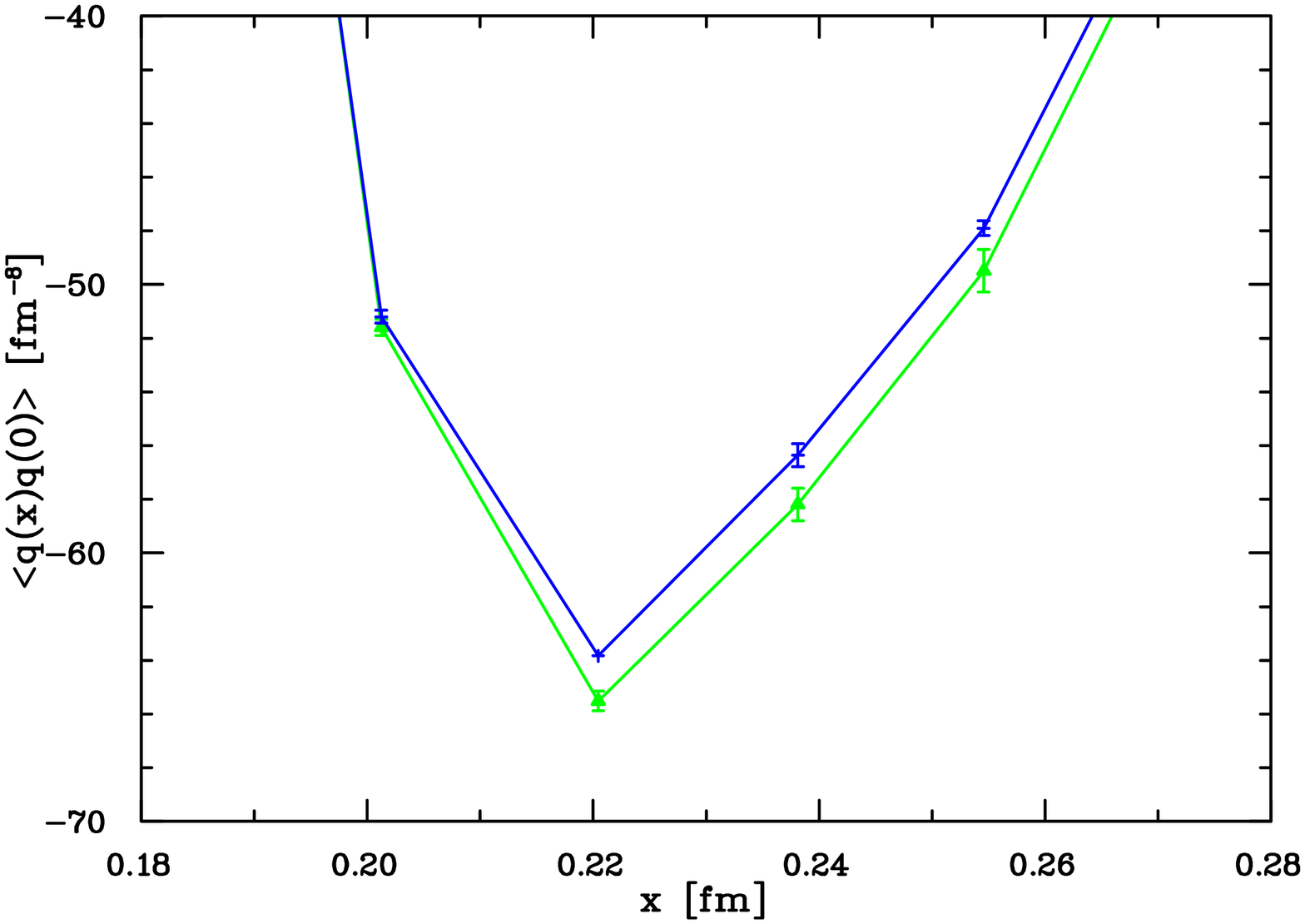}
  
  \caption{A comparison of the $\cor$ correlator for quenched and
    dynamical gauge fields. The smaller $\beta$ values used for the
    dynamical fields result in greater field fluctuations that are
    visible through an increase in the magnitude of the negative
    dip. Although not shown, the contact term $\langle q^2(0)
    \rangle_x$ has also increased and values are given in the
    text. The lower graph displays the same data as the upper graph,
    focusing on the negative dip for the dynamical fields. The
    magnitude of the dip is greater for the lighter quark mass.}
  \label{fig:qqcomparison}
\end{figure}

Given that in the chiral limit $\chi^{dyn} = 0$, it follows that an
increasing mean-square density $\langle q^2(0) \rangle_x$ must be
compensated for by a stronger negative dip in the $\cor$ correlator.
The $\cor$ correlator for the quenched ensemble and the two dynamical
ensembles is shown in\efig{fig:qqcomparison}.
These correlators were generated using four sweeps of over-improved
stout-link smearing. We see that the contact term $\langle q^2(0)
\rangle_x$ is larger and the magnitude of the negative dip has also
increased. These effects are stronger for lighter quark masses. This
is of course expected because smaller quark masses require smaller
$\beta$ values.  The exact values of the positive contact term are;
quenched $= 2924\pm 4\, {\rm fm^{-8}}$, heavy $= 5251 \pm 12 \, {\rm
  fm^{-8}}$, light $= 5432 \pm 8 \, {\rm fm^{-8}}$

Hasenfratz\ecite{Hasenfratz} also observed an increase in the negative
contribution to $\chi$ with the addition of fermion loops. However,
their correlators also show a decrease in the mean square density,
which is in contrast to our results. This may be explained through the
use of a standard smoothing algorithm, which is renowned for
destroying topological objects in the gauge field.

\section{Instanton-Like Objects}
\label{sec:instantonlikeobjects}

\subsection{Profile Versus Charge Density}

Understanding the nature of instanton-like objects in the QCD vacuum
continues to be an active of area of investigation. Considerable UV
filtering reveals the presence of long-distance topological structures
in the QCD vacuum. These topological objects are only approximations
to the classical instanton solution, but are commonly referred to as
instantons. The amount of smoothing usually applied in order to
observe these instanton-like objects is far greater than the four
sweeps of over-improved stout-link smearing that we used in the
calculation of the $\cor$ correlator.

We now analyze the similarity of the topological objects in the QCD
vacuum to the classical instanton solution.  Using over-improved
stout-link smearing we are able to extract both the action and charge
densities of our gauge fields. Starting with the action density we
locate the positions of all local maxima in the field.  The local
maxima are identified by finding a point at the center of a $3^4$
hypercube whose action density exceeds that of the neighboring 80
points of the hypercube.

Taking each maxima to be the approximate center of a possible
instanton-like object we fit the classical instanton action density
\begin{equation}
  S_0(x) = \xi \, \frac{6}{\pi^2} \, \frac{ \rho_{\rm inst}^4 }
  { \left ( (x-x_0)^2 + \rho_{\rm inst}^2 \right )^4} \, .
  \label{eqn:S0x}
\end{equation}
to the measured action density.  An arbitrary scale factor, $\xi$, is
included to allow the {\it shape} of the action density to determine
the size, $\rho_{\rm inst}$.  We fit the six parameters, $x_0$, $\xi$
and $\rho_{\rm inst}$ by fitting Eq.~(\ref{eqn:S0x}) to the action
density of the aforementioned $3^4$ hypercube.

From $\rho_{\rm inst}$, one can infer the topological charge to be
observed at the center of the distribution $q(x_0)$ if it truly is an
instanton
\begin{equation}
  q(x_0) = Q\frac{6}{\pi^2\rho_{\rm inst}^4} \,.
  \label{eqn:qx0}
\end{equation}
Here $Q = \pm 1$ for an instanton/anti-instanton.  This can then be
compared with the topological charge measured directly from the charge
density observed on the lattice.

\begin{figure}
  
  \includegraphics[angle=90,width=0.4\textwidth]{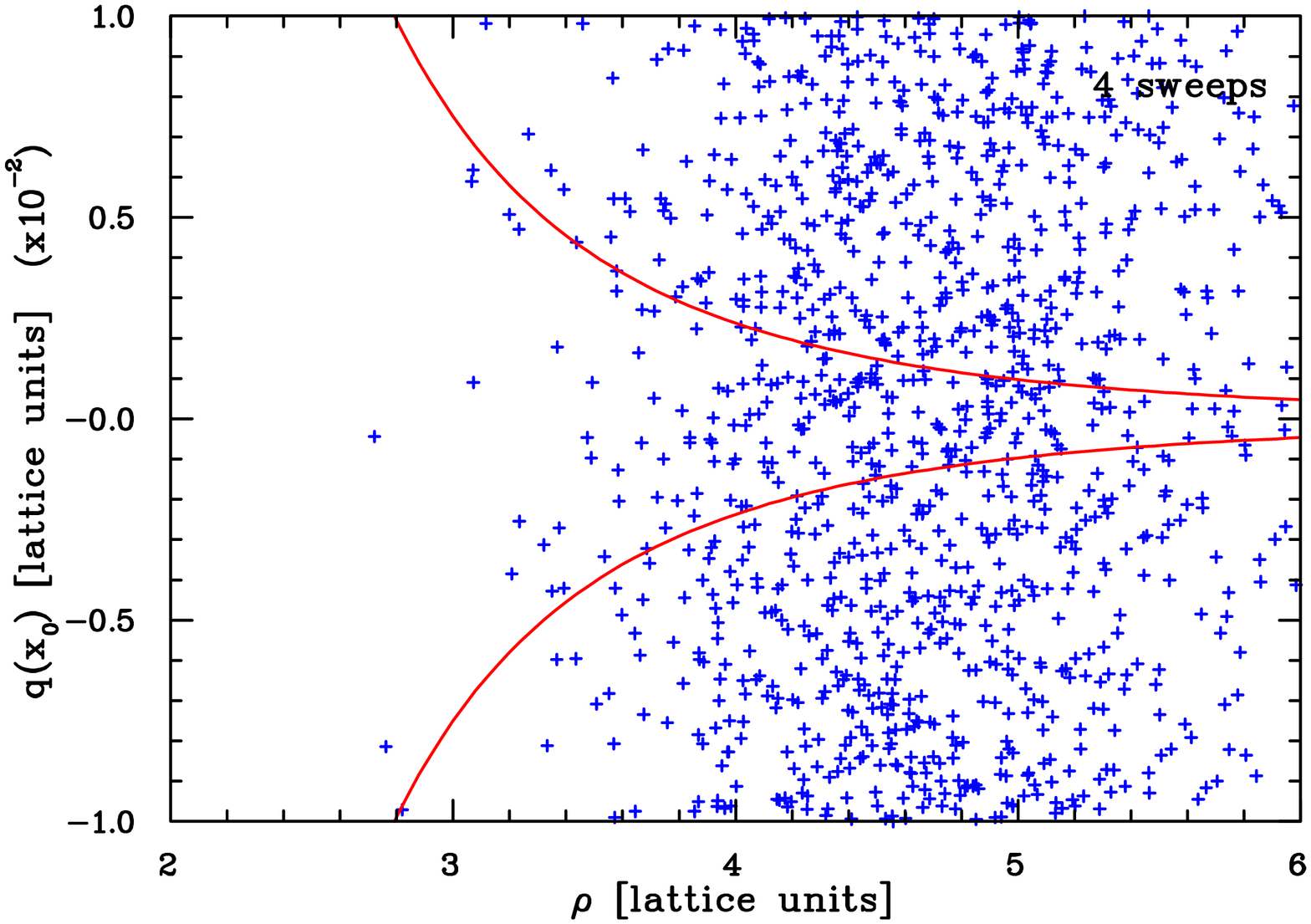}\\[10pt]
  \includegraphics[angle=90,width=0.4\textwidth]{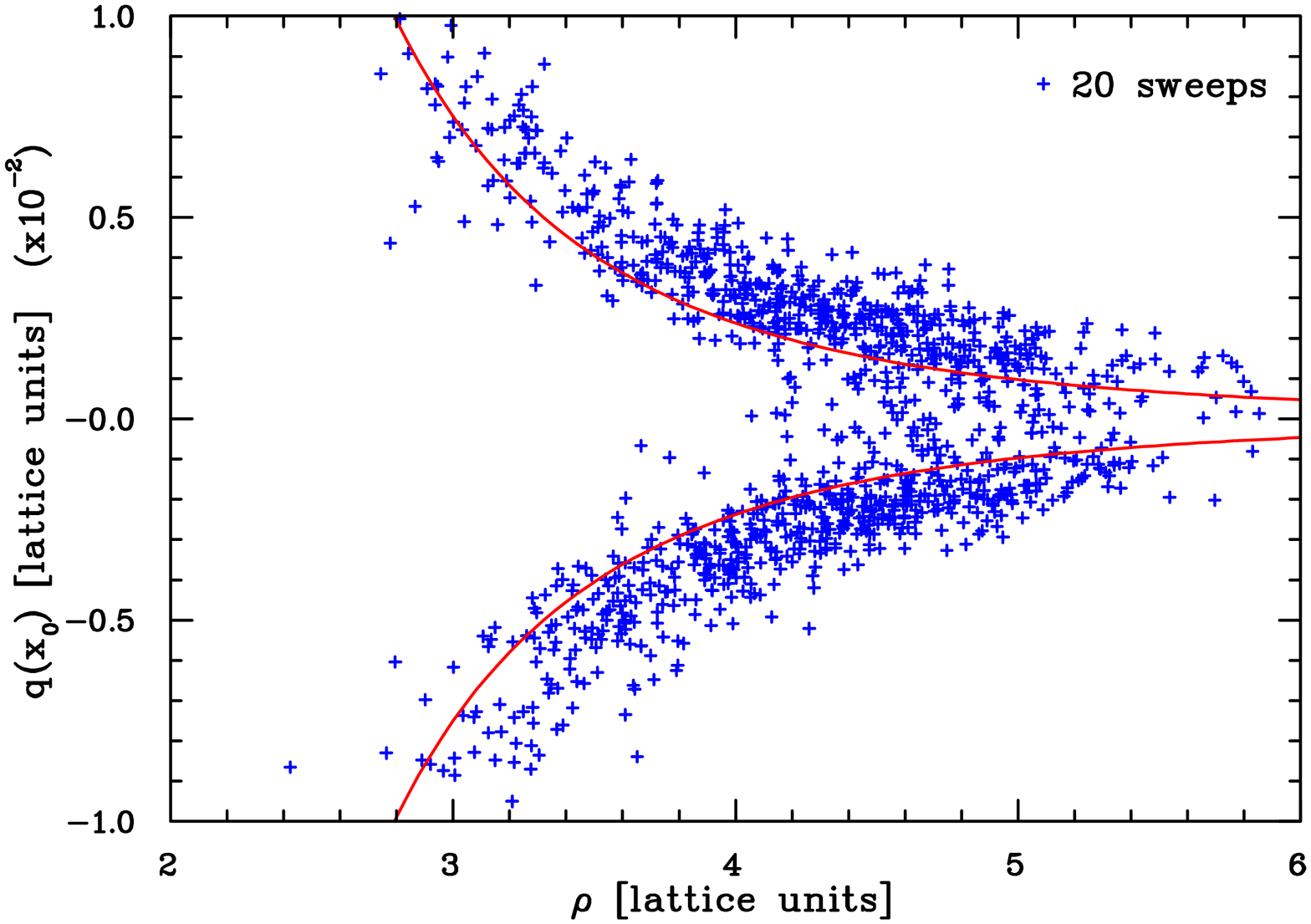}\\[10pt]
  \includegraphics[angle=90,width=0.4\textwidth]{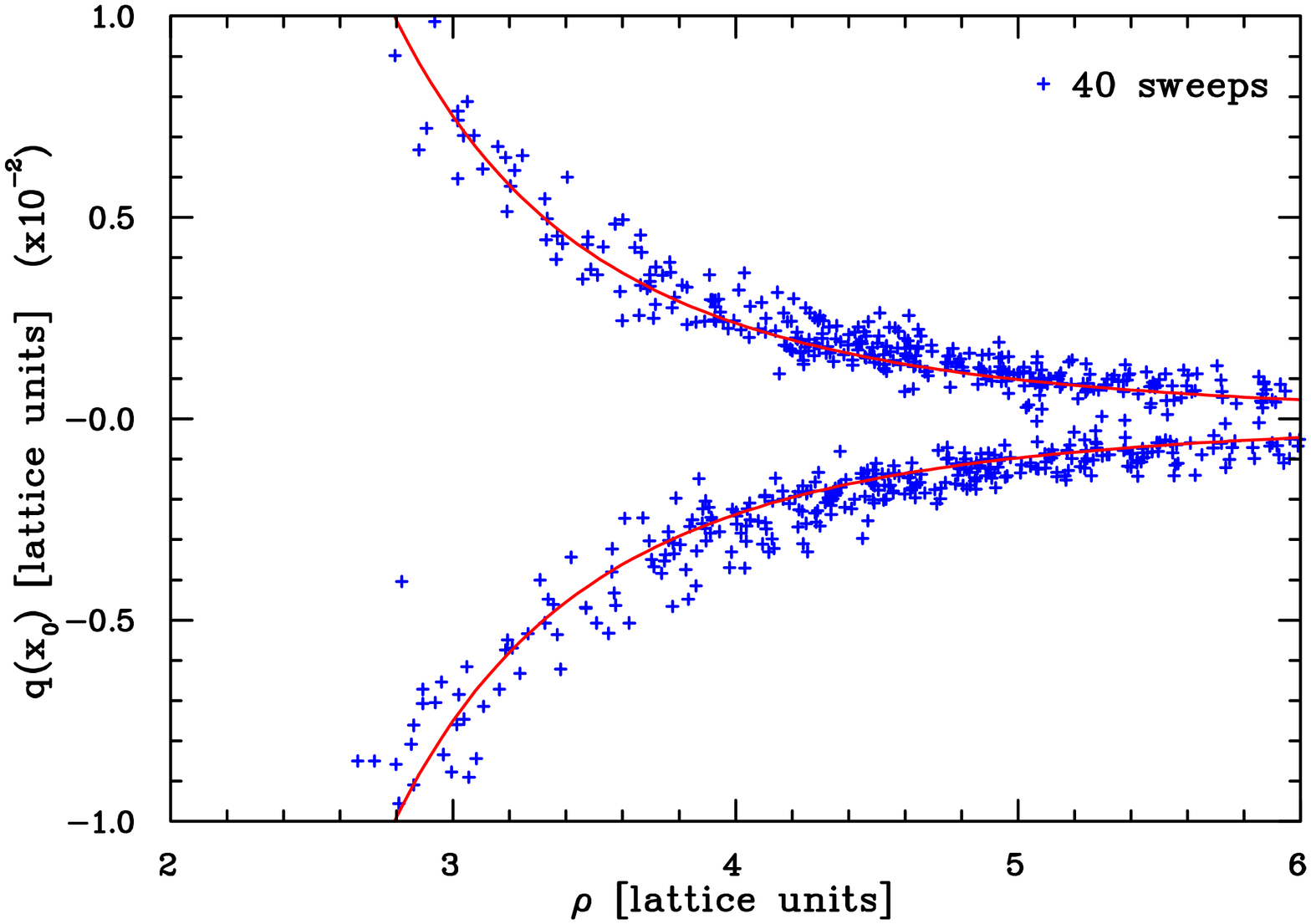}
  
  \caption{A comparison of the charge at the center of the
    instanton-like objects $q(x_0)$ observed on the lattice versus
    their size $\rho_{\rm inst}$ obtained via Eq.~(\ref{eqn:S0x}) for
    three different levels of over-improved stout-link smearing.  From
    top to bottom we apply 4 sweeps, 20 sweeps and 40 sweeps of
    smearing.  Each cross represents a local maxima in the action
    density of the gauge field.  After 40 sweeps most points lie close
    to the classical curve.}
  \label{fig:fiticomparison}
\end{figure}

A calculation on the ensemble of quenched gauge fields is presented
in\efig{fig:fiticomparison}.
Each cross represents a peak in the action density. If the peak were
to exactly represent an instanton it would lie on the theoretical
curve of Eq.~(\ref{eqn:qx0}).  For a small number of sweeps we see
that the peaks in the action density do not correspond well to
instantons. It is only after extensive smearing has been applied that
the objects start to approximate instantons.

After 40 sweeps of over-improved stout-link smearing we have a large
number of maxima in the vacuum which appear to be good approximations
to the classical instanton solution, at least within the $3^4$
hypercube used to determine the ``instanton'' size.

While the core shape and density approximate an instanton, we now
assess how similar the remainder of the object is to a classical
instanton.  To do so, we consider all points within a distance $r$
measured relative to the instanton size $\rho_{\rm inst}$ and examine
the extent to which the points within this distance have the same sign
topological-charge density as observed at the center, $q(x_0)$. If the
detected object is a good approximation to a classical instanton then
all these points should have the same charge as $q(x_0)$.

\begin{figure}
  
  \includegraphics[angle=90,width=0.45\textwidth]{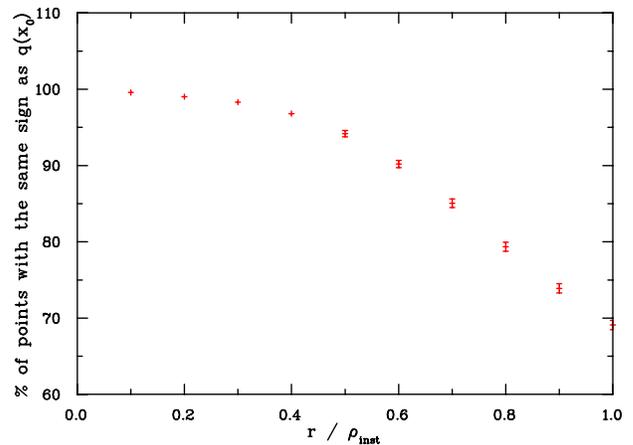}
  
  \caption{The percentage of points that are sign coherent within a
    relative distance $r/\rho_{\rm inst}$ of each instanton-like
    objects' center $x_0$ with size $\rho_{\rm inst}$.  For small
    $r/\rho_{\rm inst}$ the percentage of sign-coherent points is
    close to 100, however the sign-coherence falls off rapidly as $r$
    approaches the characteristic size $\rho_{\rm inst}$.}
  \label{fig:fiti.xonrho.ps}
\end{figure}

In Fig.~\ref{fig:fiti.xonrho.ps} we show the percentage of points that
are within a relative distance $r/\rho_{\rm inst}$ of $x_0$ that have
the same sign as $x_0$. For small $r/\rho_{\rm inst}$ the percentage
of sign-coherent points is close to 100, however it falls off rapidly
as $r$ approaches the characteristic size $\rho_{\rm inst}$.  This
suggests that although the object is representative of an instanton at
its center, the tails of the objects are distorted by vacuum
fluctuations.  What is remarkable is that at the characteristic size
of the ``instanton,'' merely 2/3 of the points are sign coherent,
suggesting that the objects revealed after 40 sweeps of smearing are
good approximations of classical instantons only at the core.

\subsection{Dynamical Fermions and Instanton Structure}
\label{subsec:dynamicalfermionsandinstantonstructure}

Given the strong correlation between $q(x_0)$ extracted from the
action density after 40 sweeps of smearing and that given by the
classical solution\eref{eqn:qx0} we can now compare ``instanton''
distributions between quenched and dynamical QCD.

\begin{figure}
  
  \includegraphics[angle=90,width=0.4\textwidth]{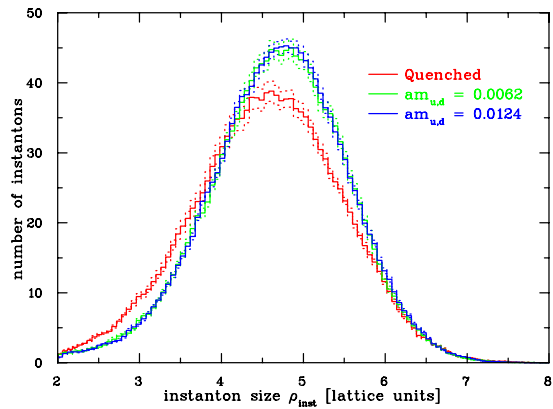}
  
  \caption{ A histogram of the instanton size $\rho_{\rm inst}$ in
    dynamical and quenched gauge fields. The dynamical fields show
    both an increase in the number of instantons detected and in
    average instanton size.}
  \label{fig:hists}
\end{figure}

We examine the variation in instanton size between the different gauge
fields. A histogram of $\rho_{\rm inst}$ is shown in\efig{fig:hists}.
Compared to the quenched ensemble, the dynamical gauge fields show an
increase in average instanton size and also in the total number of
instantons. A similar comparison was carried out in Ref.\ecite{Brower}
where it was argued that the instanton distributions were the same
between quenched and dynamical gauge fields, however the results of
that study were limited by statistical fluctuations.

It is often argued that the increased density of instanton-like
objects on the lattice can be explained through an
instanton/anti-instanton attraction occurring due to the presence of
the fermion determinant in the QCD weight
factor~\cite{Hart:2001pj}. Isolated instantons and anti-instantons
give rise to zero modes of the Dirac operator. When generating
dynamical gauge fields the selection of typical configurations is
weighted by $\det (\slashed{D}+m) e^{-S_g}$.  If an exact zero-mode of
$\slashed{D}$ were to exist on the lattice then the determinant would
approach $0$ in the chiral limit and it would be highly improbable
that the configuration would be selected. Thus, isolated instantons
will not exist in the light quark mass gauge fields and hence all
instanton-like objects will be closer in these fields.  This naturally
leads one to expect that there will be a greater number of instantons
in the dynamical gauge fields. However, this does not account for the
increased average size of the topological objects, which is an
interesting result.

\section{Topological Charge Density}
\label{sec:topologicalchargedensity}

The effects of dynamical quarks are realized significantly in the
calculation of the $\cor$ correlator. The increased magnitude of the
non-trivial topological charge field fluctuations that are permitted
due to the inclusion of fermion loops should also be visible in direct
visualizations of the topological charge density.

Using four sweeps of over-improved stout-link smearing, we consider
the short-range structure of the topological charge density.
\begin{figure*}
  \includegraphics[width=0.9\textwidth]{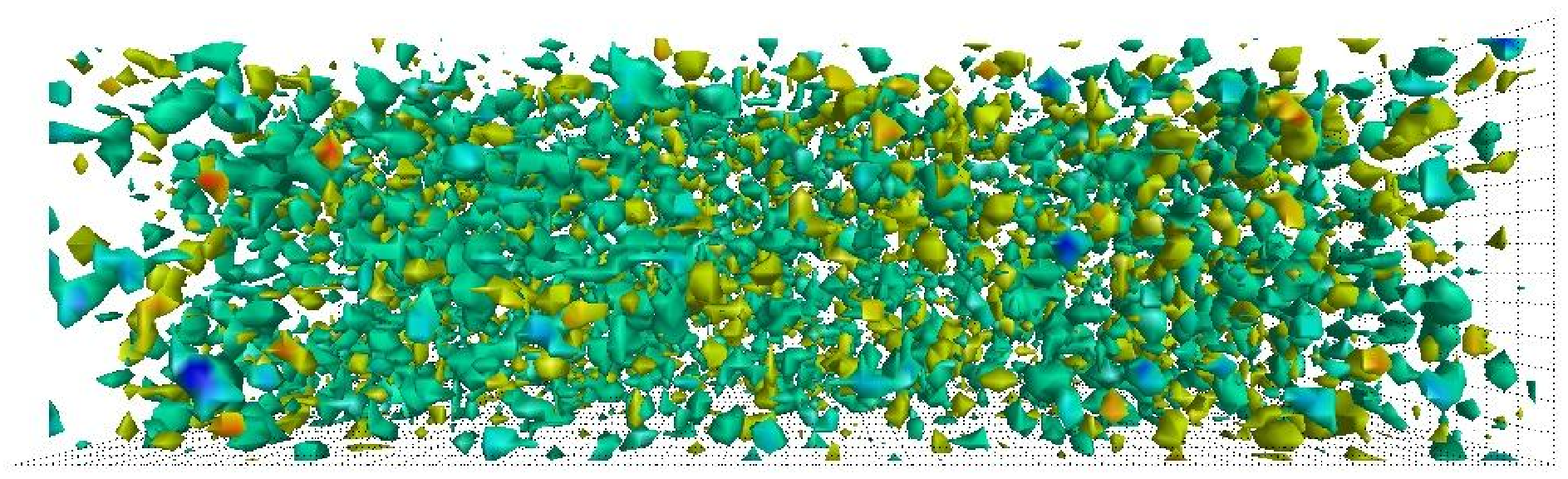}\\[10pt]
  \includegraphics[width=0.9\textwidth]{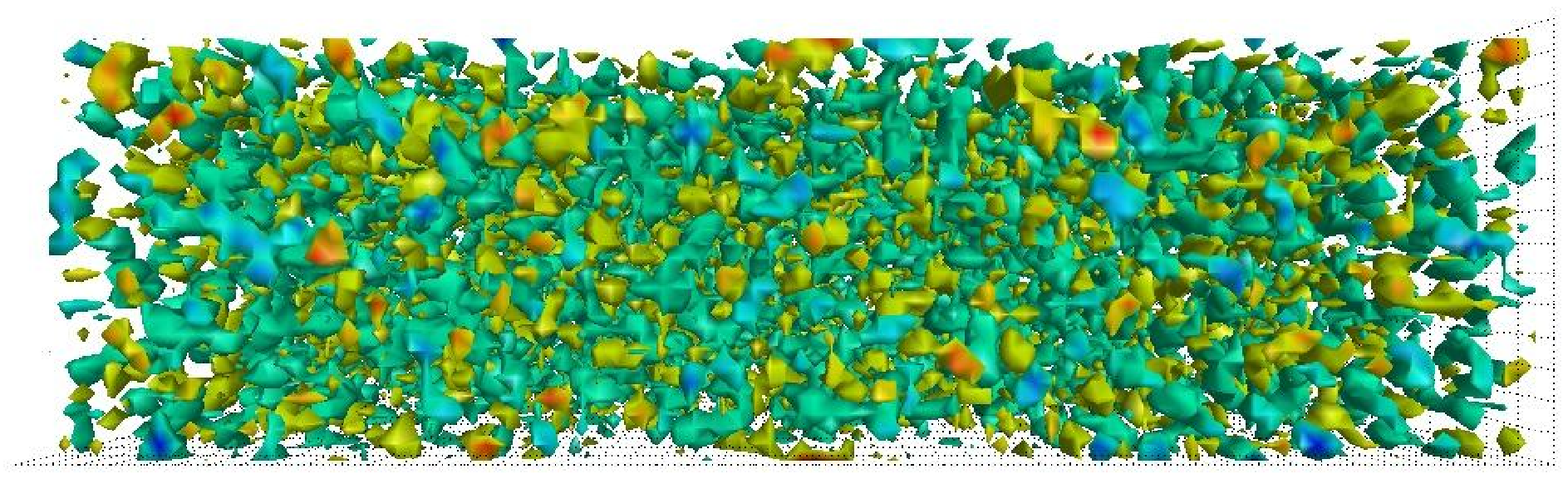}\\[10pt]
  \includegraphics[width=0.9\textwidth]{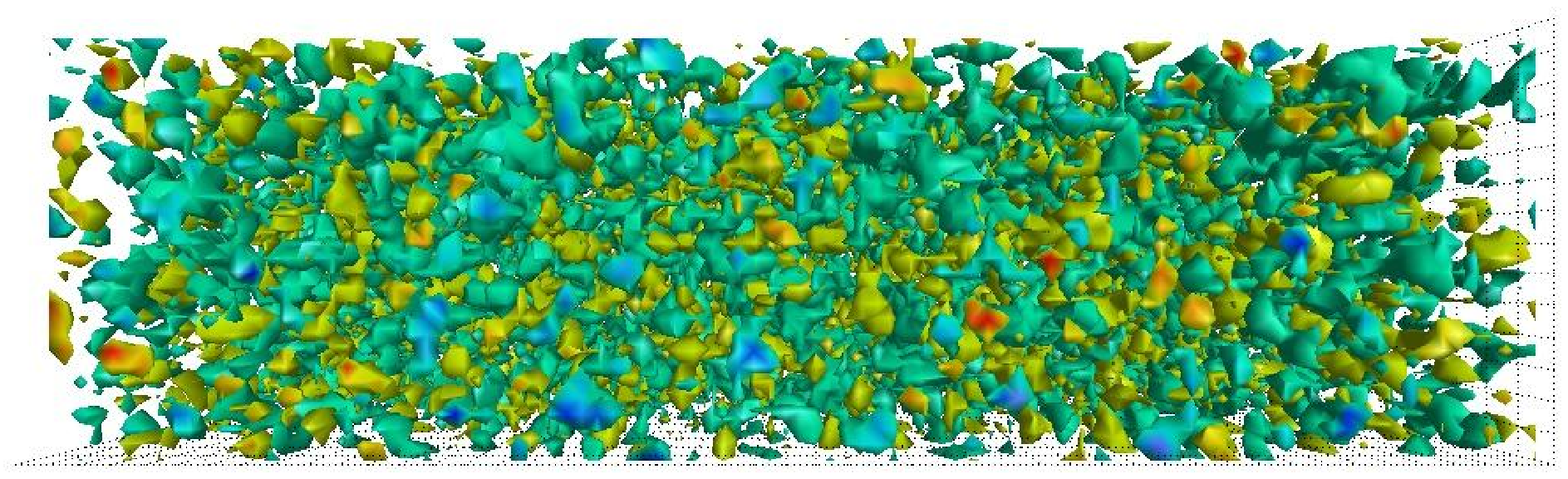}
  \caption{The topological charge density for the quenched and
    dynamical ensembles, obtained after four sweeps of over-improved
    stout-link smearing. From top to bottom we plot a quenched field,
    the dynamical $am_{u,d} = 0.0124$, $am_s = 0.031$ field and the
    dynamical $am_{u,d} = 0.0062$, $am_s = 0.031$ field. The greater
    non-trivial field excitations that are permitted upon the
    introduction of light dynamical fermions are directly visible in
    the dynamical illustrations.}
  \label{fig:topqx}
\end{figure*}
In\efig{fig:topqx} we present the topological charge density for the
quenched and two dynamical ensembles. The extra field fluctuations
that are permitted due to the fermion determinant are clearly visible
in the visualizations of the dynamical QCD vacuum.

\begin{figure*}
  \includegraphics[width=0.9\textwidth]{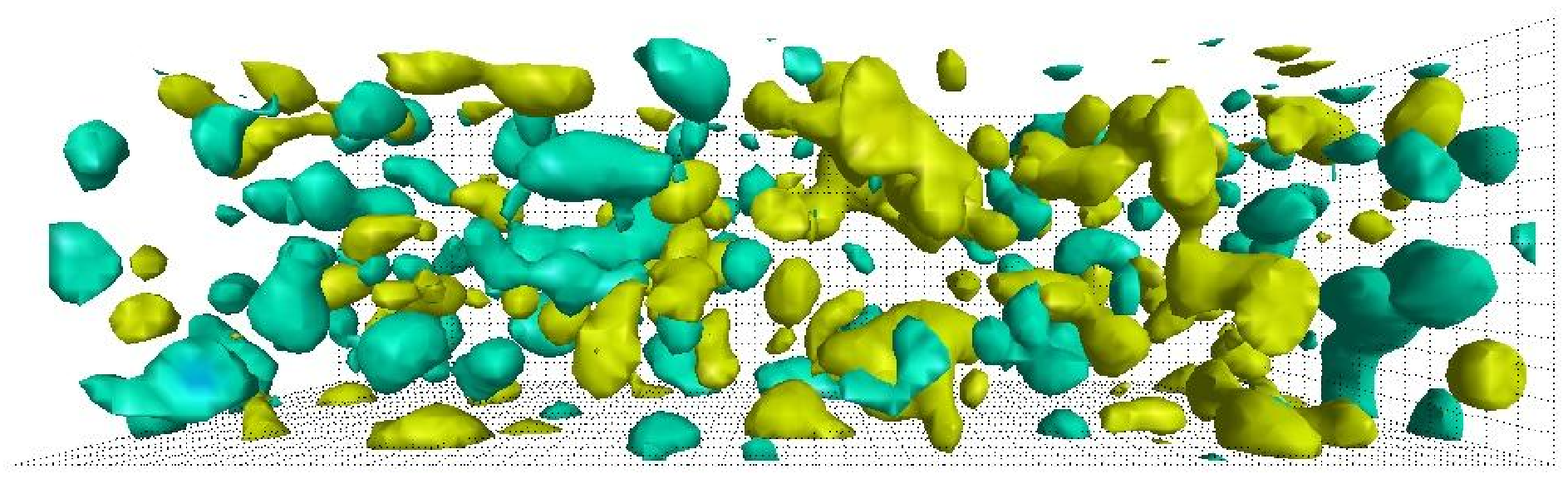}\\[10pt]
  \includegraphics[width=0.9\textwidth]{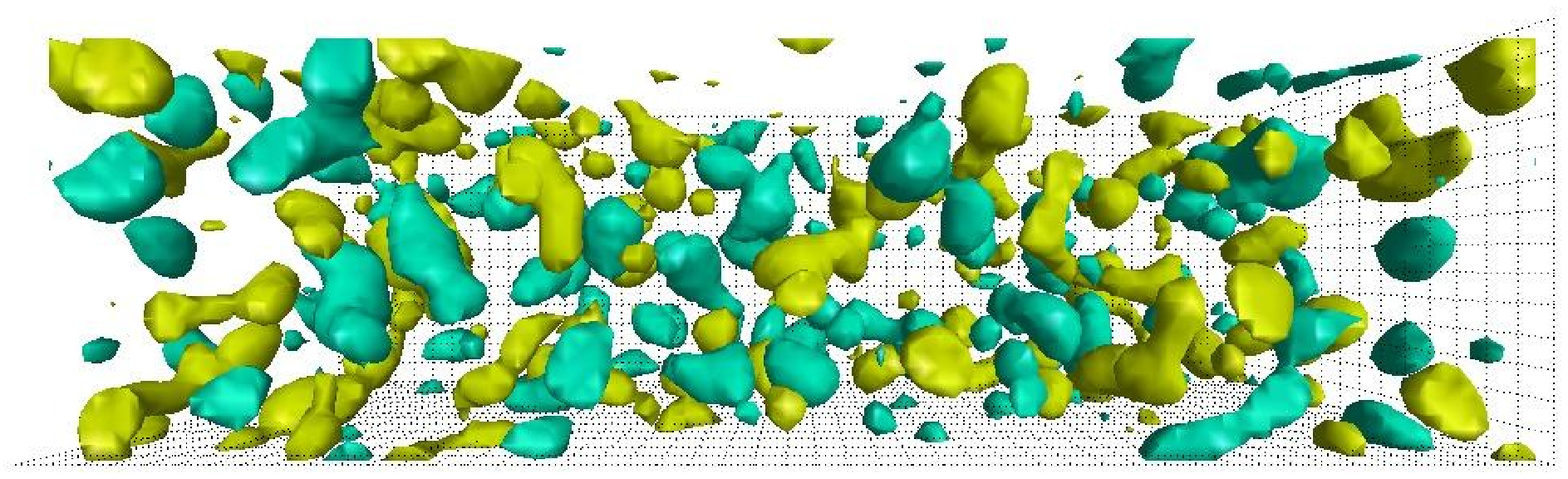}\\[10pt]
  \includegraphics[width=0.9\textwidth]{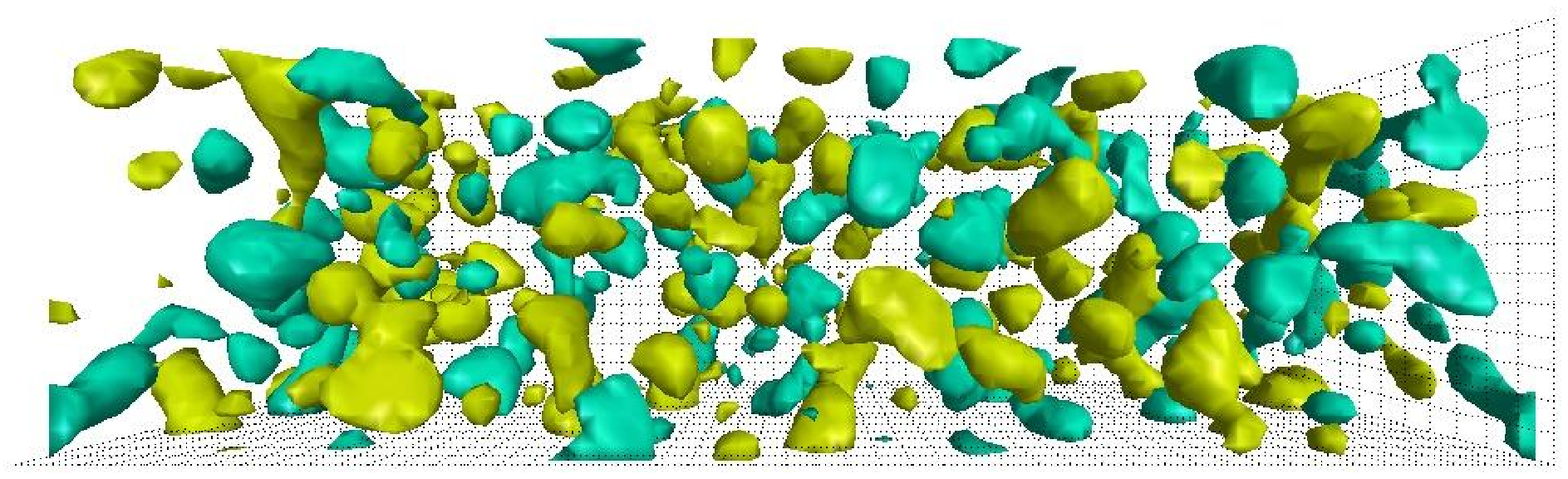}
  \caption{The topological charge density for the quenched and
    dynamical ensembles, obtained after forty sweeps of over-improved
    stout-link smearing. From top to bottom we plot a quenched field,
    the dynamical $am_{u,d} = 0.0124$, $am_s = 0.031$ field and the
    dynamical $am_{u,d} = 0.0062$, $am_s = 0.031$ field. It is
    difficult to observe a noticeable difference in the
    size-distribution of the gauge field fluctuations when observing a
    single time-slice, but the sparseness of the quenched field is
    readily observed.}
  \label{fig:smoothtopqx}
\end{figure*}

In\efig{fig:smoothtopqx} we compare the structure of the vacuum after
40 sweeps of over-improved smearing.
It is difficult to observe the increased density of instantons in
these figures. This is because the charge density fluctuates over
different time-slices and a single time-slice does not always portray
an accurate representation of the vacuum.

\section{Conclusion}
\label{sec:conclusion}

By applying two different levels of smearing to the large-volume MILC
gauge fields, we explored the effects of dynamical fermions on
topology.  The results agree with expectations.  The addition of
fermions into the QCD action at constant lattice spacing renormalizes
the coupling constant such that the coupling parameter $\beta$ becomes
smaller. This permits greater field fluctuations and is manifest
through an increased RMS density of topologically non-trivial field
fluctuations.  This induces an increased negative dip in the
topological charge density correlator and a larger contact term
reflecting the RMS density of topological charge density.  This effect
increases as $m_{u,d} \rightarrow 0$.

The results reflect the suppression of zero-modes due to the inclusion
of the $\det(\slashed{D}+m)$ weight factor in the selection of typical
gauge fields, resulting in a decrease in the number of isolated
instanton-like objects. This causes instantons and anti-instantons to
be ``attracted''~\cite{Hart:2001pj} and leads to an increase in the
density of instanton-like objects in the dynamical gauge fields.

These results support the emerging picture of the vacuum as an
alternating ``sandwich'' of opposite topological charge density.
Beneath this oscillating short-range structure there exists a
long-distance foundation of instanton-like objects that can be
revealed through smoothing.  The addition of dynamical fermions allows
stronger field fluctuations and a higher frequency of
sign-oscillations in the topological charge density. The density of
instanton-like objects beneath these short-distance oscillations also
increases, as does their average size.

\section*{Acknowledgments}
\label{sec:acknowledgements}

The authors thank Ernst-Michael Ilgenfritz and Waseem Kamleh for
constructive and insightful discussions.  We also thank the Australian
Partnership for Advanced Computing (APAC) and the South Australian
Partnership for Advanced Computing (SAPAC) for generous grants of
supercomputer time which have enabled this project. This work is
supported by the Australian Research Council.

\end{document}